\documentclass[letter,twocolumn]{jpsj3}
\newcommand{\vtr}[1]{
\mbox{\boldmath $ #1 $}}

\usepackage{bm}
\usepackage{graphicx}
\usepackage{color}

\def\runauthor{Online-Journal Subcommittee}

\title{%
An Efficient Monte-Carlo Method for Calculating Free Energy in Long-Range Interacting Systems\\
}

\author{
Kazuya \textsc{Watanabe} and Munetaka \textsc{Sasaki}\thanks{E-mail : msasaki@camp.apph.tohoku.ac.jp}
}

\inst{
Department of Applied Physics, Tohoku University, Sendai 980-8579
}

\recdate{\today}

\abst{We present an efficient Monte-Carlo method for long-range interacting systems 
to calculate free energy as a function of an order parameter. 
In this method, a variant of the Wang-Landau method regarding the order parameter 
is combined with the stochastic cutoff method, which has recently been developed for long-range 
interacting systems. This method enables us to calculate free energy in long-range interacting
systems with reasonable computational time despite the fact that no approximation is involved. 
This method is applied to a three-dimensional magnetic dipolar system
to measure free energy as a function of magnetization. 
By using the present method, we can calculate free energy for 
a large system size of $16^3$ spins despite the presence of long-range magnetic dipolar interactions.
We also discuss the merits and demerits of the present method in comparison with 
the conventional Wang-Landau method in which free energy is calculated from 
the joint density of states of energy and order parameter.}

\kword{
Monte Carlo, long-range interacting system, Wang-Landau method, free energy measurement, 
magnetic dipolar system
}

\begin{document}
\maketitle


In general, Monte Carlo (MC) simulations in long-range interacting systems are much more 
difficult than those in short-range interacting systems because we have to 
take a large number of interactions 
into consideration. For example, in the case of systems with pairwise interactions, 
the number of interactions is proportional to $N^2$, where $N$ is the number of elements 
of the system. Therefore, if a naive MC simulation is carried out in such systems, 
the computational time per MC step rapidly increases in proportion 
to $N^2$, which is in contrast to that in the case of short-range 
interacting systems in which the computational time increases
in proportion to $N$. In order to overcome this difficulty, 
many simulation methods have been proposed~\cite{Appel85,Barnes86,Greengard88,Carrier88,Saito92,Ding92,LuijtenBlote95,Sasaki96,
Hetenyi02, Bernacki04, FukuiTodo09}. 

Recently, one of the authors and Matsubara have developed an efficient 
MC method called the stochastic cutoff (SCO) method 
for long-range interacting systems~\cite{SasakiMatsubara08}. In the SCO method, 
each of the pairwise interactions $V_{ij}$ is stochastically switched to either $0$ 
or a pseudointeraction ${\bar V}_{ij}$ 
by the stochastic potential switching (SPS) algorithm~\cite{Mak05, MakSharma07}. 
The switching probability to $0$ and that to ${\bar V}_{ij}$ are 
$P_{ij}$ and $1-P_{ij}$, respectively. 
Since the pseudointeraction ${\bar V}_{ij}$ and switching probability $P_{ij}$ are chosen 
properly in the SPS algorithm, the SCO method strictly satisfies the detailed balance condition 
concerning the original Hamiltonian~\cite{Mak05, MakSharma07, Sasaki10}. This means that the SCO method does not 
involve any approximation. Furthermore, since most of the distant and weak interactions are switched to $0$ 
and an efficient method to switch potentials has been developed~\cite{SasakiMatsubara08}, 
the SCO method enables us to reduce the computational time of long-range interactions markedly. 
For example, in the case of three-dimensional dipolar systems, to which our new method will be applied later, 
the computational time is reduced from ${\cal O}(N^2)$ to ${\cal O}(N \log N)$ by the SCO 
method~\cite{SasakiMatsubara08}. 
We can measure internal energy and heat capacity without ${\cal O}(N^2)$ computation 
by a method proposed in ref.~\citen{Sasaki10}. An efficient method to combine the SCO method 
with the replica exchange method~\cite{HukushimaNemoto96} has also been developed there. 

In this letter, we propose an efficient MC method of combining the SCO method with the Wang-Landau 
method~\cite{WangLandau01A, WangLandau01B}. This method enables us to calculate free energy 
as a function of an order parameter with reasonable computational time even in long-range 
interacting systems. In the case of three-dimensional magnetic 
dipolar systems, computational time per MC step is reduced 
from ${\cal O}(N^2)$ to ${\cal O}(N\log N)$. 
As will be shown later, this method enables us to calculate free energy 
in a three-dimensional magnetic dipolar system with a size of $16^3$ spins.
This system size is much larger than that of the previous work~\cite{Zhou06}, {\it i.e.}, 
$10^3$ spins, in spite of the fact that long-range dipolar interactions are included 
in the present work and they are not included in the previous work. 


We now start to present our new MC method. As an example, we hereafter consider to measure free energy 
as a function of the z-component of the magnetization in a classical Heisenberg spin system. It is straightforward 
to generalize the method for other cases. Free energy is defined by
\begin{eqnarray}
&&\exp[-\beta F(\beta;m_z)] \nonumber \\
&&\equiv C {\rm Tr}_{\{\vtr{S}_i\}} \exp[-\beta {\cal H}\{\vtr{S}_i\}]\delta(m_z-m_z^*\{\vtr{S}_i\}),
\label{eqn:FeneDef}
\end{eqnarray}
where 
\begin{equation}
m_z^*\{\vtr{S}_i\}\equiv \frac1N \sum\nolimits_{i}S_i^z,
\label{eqn:MzDef}
\end{equation}
where $N$ is the number of spins, $\beta$ is the inverse temperature, ${\cal H}\{\vtr{S}_i\}$ is the Hamiltonian 
of the spin system, and $C$ is a constant. It is not important how we choose $C$ 
because it only contributes to $F(\beta;m_z)$ as a constant. 
The right-hand side of eq.~(\ref{eqn:FeneDef}) is 
the sum of the weights of all the states with a magnetization $m_z$.

In the present method, we use a variant of the Wang-Landau method regarding the order parameter. 
A similar method has been proposed in ref.~\citen{Berg93} to calculate free energy 
by the multicanonical ensemble method~\cite{BergNeuhaus91,BergNeuhaus92}. 
The basic idea is as follows: We perform an MC simulation 
with the Hamiltonian
\begin{equation}
{\cal H}'\{ \vtr{S}_i \} = {\cal H}\{ \vtr{S}_i \}-\beta^{-1}G(m_z^*\{\vtr{S}_i\}).
\label{eqn:Hamiltonian2}
\end{equation}
During the simulation, $G(m_z)$ in eq.~(\ref{eqn:Hamiltonian2}) is modified so that 
$P(m_z)$ becomes a constant, where $P(m_z)$ is the probability that 
a state with a magnetization $m_z$ is sampled. Then, the resultant function $G(m_z)$ 
is related to the free energy $F(\beta;m_z)$ by
\begin{equation}
G(m_z)=\beta F(\beta;m_z) + {\rm constant}.
\end{equation}
This can be easily shown as
\begin{eqnarray}
&&\hspace*{-8mm}P(m_z) \nonumber \\
&&\hspace*{-8mm}\propto {\rm Tr}_{\{\vtr{S}_i\}} \exp[-\beta {\cal H}'\{\vtr{S}_i\}]
\delta(m_z-m_z^*\{\vtr{S}_i\}) \nonumber \\
&&\hspace*{-8mm}= \exp[G(m_z)] {\rm Tr}_{\{\vtr{S}_i\}} \exp[-\beta {\cal H}\{\vtr{S}_i\}]
\delta(m_z-m_z^*\{\vtr{S}_i\}) \nonumber \\
&&\hspace*{-8mm}= \exp[G(m_z)] C^{-1} \exp[-\beta F(\beta;m_z)] = {\rm constant},
\end{eqnarray}
where we have used eq.~(\ref{eqn:FeneDef}) to go from the third line 
to the fourth. To modify $G(m_z)$ so that $P(m_z)$ becomes a constant, 
we use the conventional procedure of the Wang-Landau method~\cite{WangLandau01A,WangLandau01B}, 
{\it i.e.}, we modify $G(m_z)$ as 
\begin{equation}
G(m_z) \rightarrow G(m_z) - \Delta F,\quad(\Delta F > 0)
\label{eqn:WLprocedure}
\end{equation}
after each trial to update a single spin. 
If we start our simulation with $G(m_z)=0$, as conventionally performed, states with low 
free energies are frequently sampled at the beginning of the simulation. 
However, since the weights of such states are reduced more by the reduction in $G(m_z)$ 
[recall that the Hamiltonian is given by eq.~(\ref{eqn:Hamiltonian2})], 
$G(m_z)$ is adjusted by this procedure so that $P(m_z)$ becomes a constant. 
As is conventionally performed in the Wang-Landau method~\cite{WangLandau01A, WangLandau01B}, the constant $\Delta F$ 
in eq.~(\ref{eqn:WLprocedure}) is gradually reduced as the simulation proceeds. 

Now, we briefly compare the present method with the conventional Wang-Landau method 
in which one evaluates the joint density of states as a function of energy and magnetization
\begin{equation}
n(E,m_z)\equiv {\rm Tr}_{\{\vtr{S}_i\}} \delta(E-{\cal H}\{\vtr{S}_i\})\delta(m_z-m_z^*\{\vtr{S}_i\}),
\label{eqn:jointDOS}
\end{equation}
and calculates free energy from $n(E,m_z)$ as
\begin{equation}
\exp[-\beta F(\beta;m_z)] = \int_{-\infty}^{\infty} {\rm d}E \exp(-\beta E) n(E,m_z).
\label{eqn:FEintegral}
\end{equation}
In the conventional Wang-Landau method, 
we modify a two-variable function $G(E,m_z)$, which becomes proportional to 
$n(E,m_z)$ at the end of the simulation, by a method similar to eq.~(\ref{eqn:WLprocedure}).
Therefore, we have to calculate $E$ and $m_z$ after each update of a single spin. 
Since the number of interactions per spin is $N-1$ in long-range interacting 
systems, the computational time for calculating the new energy is ${\cal O}(N)$. 
In contrast, we only need to calculate $m_z$ in the present method, 
and the computational time to calculate the new magnetization is ${\cal O}(1)$. 
This is the reason why we use the variant of the Wang-Landau method.

Note that the above-mentioned simulation method is still time-consuming 
if ${\cal H}$ in eq.~(\ref{eqn:Hamiltonian2}) involves long-range interactions
because we need to carry out an MC simulation with the Hamiltonian ${\cal H}'$. 
To overcome such difficulty, we simply use the SCO method. Since the SCO method with the Hamiltonian ${\cal H}'$ 
samples a state according to the Boltzmann weight $\exp(-\beta {\cal H}')$ 
as in the conventional MC method, we can measure $F(\beta;m_z)$ in the same way as before. 
In the case of three-dimensional magnetic dipolar systems, 
the number of interactions per spin is reduced from $N-1$ to ${\cal O}(\log N)$ 
by the SCO method~\cite{SasakiMatsubara08}. 
Therefore, the computational time per MC step becomes ${\cal O}(N \log N)$. 
Because the SCO method can be applied to a part of the Hamiltonian~\cite{Mak05, Sasaki10}, 
we only apply the SCO method to a long-range part in ${\cal H}$. 

To summarize, we show the whole procedure of our method. 
When the original Hamiltonian consists of long-range interactions 
${\cal H}^{\rm (L)}\{ \vtr{S}_i \}=\sum_{i<j}V_{ij}(\vtr{S}_i,\vtr{S}_j)$ 
($V_{ij}$ is a pairwise interaction) 
and short-range ones ${\cal H}^{\rm (S)}\{ \vtr{S}_i \}$, the procedure proceeds as follows: 
\begin{itemize}
\item[1)] Set $G(m_z)=0$ and $H(m_z)=0$, where $H(m_z)$ is a histogram to check whether or not 
all the magnetizations are sampled with equal probabilities. The initial 
$\Delta F$ in eq.~(\ref{eqn:WLprocedure}) is set sufficiently large so that $G(m_z)$ is adjusted quickly 
in the early stage of the Wang-Landau method. 
\item[2)] Repeat the following two steps as a basic MC procedure:
\begin{itemize}
\item[a)] Switch each of $V_{ij}$ in ${\cal H}^{\rm (L)}$ to either $0$ or ${\bar V}_{ij}$ 
with a probability of $P_{ij}$ or $1-P_{ij}$, respectively. The method proposed in 
ref.~\citen{SasakiMatsubara08} is used to switch potentials efficiently. The probability $P_{ij}$ is 
\begin{equation}
P_{ij}(\vtr{S}_i,\vtr{S}_j)=\exp[\beta (V_{ij}(\vtr{S}_i,\vtr{S}_j)-V_{ij}^*)],
\label{eqn:SPSprob}
\end{equation}
where $\beta$ is the inverse temperature and $V_{ij}^*$ is a constant equal to (or greater than) 
the maximum $V_{ij}(\vtr{S}_i,\vtr{S}_j)$. The pseudopotential ${\bar V}_{ij}$ is defined by
\begin{equation}
{\bar V}_{ij}(\vtr{S}_i,\vtr{S}_j)\equiv V_{ij}(\vtr{S}_i,\vtr{S}_j)-\beta^{-1}\log[1-P_{ij}(\vtr{S}_i,\vtr{S}_j)].
\end{equation}
This potential switching step is performed every $t_{\rm switch}$ MC steps. 
\item[b)] Perform a standard MC simulation with the Hamiltonian
\begin{equation}
{\cal H}'\{ \vtr{S}_i \}={\cal H}^{\rm (S)}\{ \vtr{S}_i \}
+\sum\nolimits'{\bar V}_{ij}(\vtr{S}_i,\vtr{S}_j)
-G(m_z^*\{\vtr{S}_i\}),
\end{equation}
where the sum $\sum\nolimits'$ runs over potentials which are switched to ${\bar V}_{ij}$ in step a). 
During the simulation, $G(m_z)$ is changed according to eq.~(\ref{eqn:WLprocedure})
after each trial to update a spin. This adjustment of $G(m_z)$ is performed regardless of 
whether or not the trial is accepted. We also change the histogram as 
\begin{equation}
H(m_z)\rightarrow H(m_z)+1,
\end{equation}
after each trial. 
\end{itemize}
\item[3)] Check whether or not the histogram $H(m_z)$ is flat. If it is flat, 
halve $\Delta F$ and reinitialize the histogram as $H(m_z)=0$. 
This check is performed every $t_{\rm check}$ MC steps. 
\item[4)] Stop the simulation if $\Delta F$ is small enough. Otherwise, 
return to 2). 
\end{itemize}

To check the efficiency of the method, we apply it 
to measure free energy as a function of $m_z$ in a three-dimensional 
magnetic dipolar system. The Hamiltonian is given as
\begin{eqnarray}
&&\hspace*{-12mm}{\cal H}=-J\sum\nolimits_{\langle ij \rangle} \vtr{S}_i\cdot\vtr{S}_j
-C_{\rm u}\sum\nolimits_i(S_i^z)^2\nonumber\\
&&\hspace*{-10mm}\qquad+D\sum\nolimits_{i<j} \left[
\frac{\vtr{S}_i\cdot\vtr{S}_j}{r_{ij}^3}
-3\frac{(\vtr{S}_i\cdot\vtr{r}_{ij})(\vtr{S}_j\cdot\vtr{r}_{ij})}{r_{ij}^5}\right],
\label{eqn:TestHamiltonian}
\end{eqnarray}
where $\vtr{S}_i$ is a classical Heisenberg spin of $|\vtr{S}_i|=1$, 
$\langle ij \rangle$ runs over all the nearest-neighboring pairs, 
$\vtr{r}_{ij}$ is a vector spanned from site $i$ to site $j$ in the unit of 
the lattice constant $a$, and $r_{ij}=|\vtr{r}_{ij}|$. 
On the right-hand side of eq.~(\ref{eqn:TestHamiltonian}), the first term describes 
ferromagnetic exchange interactions, the second term uniaxial magnetocrystalline 
anisotropy energies whose easy axis is parallel to the $z$-direction, 
and the third term magnetic dipolar interactions. 
As mentioned above, the system size $N$ is $16^3$. The boundary condition is open 
in all directions. We fix the ratio $C_{\rm u}/J$ to $0.1$
and change $D/J$ to see how the structure of free energy depends 
on the strength of dipolar interactions.

\begin{figure}[t]
\begin{center}
\includegraphics[width=8.5cm]{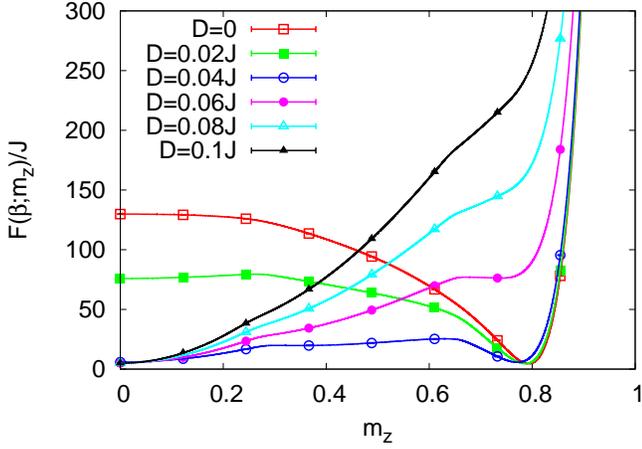}
\end{center}
\caption{(Color online) $m_z$ dependences of $F(\beta;m_z)$ for 
$D=0$, $0.02$, $0.04$, $0.06$, $0.08$, and $0.1~J$. 
The system size $N$ is $16^3$ and the temperature $T$ is $0.7~J$. All the data are shown by lines without error bars. 
Symbols are drawn with error bars at several data points. 
The average is taken over 10 different runs.
}
\label{fig:Fenergy}
\end{figure}

In the present simulations, we set the initial $\Delta F$ in eq.~(\ref{eqn:WLprocedure}) 
to $J$. We stop our simulation after we halve $\Delta F$ 20 times. 
Therefore, the final $\Delta F$ is $J\times 2^{-20}$. 
We have checked that $G(m_z)$ converges well in later stages of the Wang-Landau method. 
The histogram $H(m_z)$ is checked every 10,000 MC steps. We regard the histogram as flat 
when $H(m_z)$ for all the magnetizations is not less than 80\% of 
the average histogram $\langle H(m_z) \rangle$. 
We estimate $F(\beta;m_z)$ in the range of $0 \le m_z \le 0.98$ on a grid of $40,142$ bins. 
Since $F(\beta;m_z)$ is an even function of $m_z$, we calculate $F(\beta;m_z)$ 
for positive magnetizations. 
The SCO method is applied only to dipolar interactions. They are switched every $10$ MC steps. 
$V_{ij}^*$ in eq.~(\ref{eqn:SPSprob}) is set to the maximum $V_{ij}(\vtr{S}_i,\vtr{S}_j)$. 


The result of the free energy measurement at $T=0.7~J$ is shown in Fig.~\ref{fig:Fenergy}, 
where we set the Boltzmann constant $k_{\rm B}$ to unity. 
The temperature is well below the critical temperature of the model for $C_{\rm u}=D=0$, 
which is estimated to be about $1.44~J$\cite{Chen93}. 
For each $D$, we carried out 10 different runs with different initial conditions and 
random sequences to estimate the means and error bars of the data. 
As a result, we have found that the error bars are less than $1.0~J$ for all the data. 
In Fig.~\ref{fig:Fenergy}, symbols are drawn with error bars. The error bars 
are much smaller than the symbols. 
From the smallness of the error bars, we consider that 
correct data are obtained by the present method. 
We also see that the position of the global minimum changes 
from $m_z\approx 0.8$ to $m_z=0$ as $D$ increases. This result is reasonable 
because dipolar interactions prefer demagnetized states. The case $D=0.04~J$ is a 
marginal one in which free energies at the two minima are almost the same. 

\begin{figure}[t]
\begin{center}
\includegraphics[width=8.5cm]{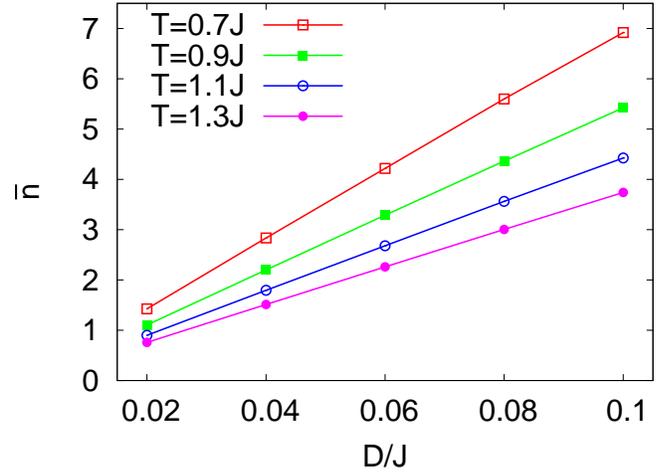}
\end{center}
\caption{(Color online) Average number ${\bar n}$ of potentials 
per site that are switched to ${\bar V}_{ij}\ne 0$ is plotted as 
a function of $D$ for $T=0.7$, $0.9$, $1.1$, and $1.3~J$ 
(from top to bottom). 
}
\label{fig:Psurvive}
\end{figure}

\begin{figure}[t]
\begin{center}
\includegraphics[width=8.5cm]{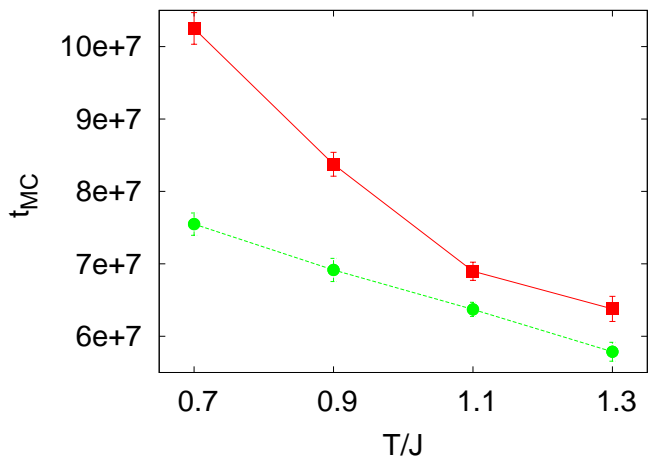}
\caption{(Color online) Temperature dependences of the total number of MC steps $t_{\rm MC}$ in the free energy 
calculation for $D=0$ (full circle) and $D=0.04~J$ (full square). 
The average is taken over 50 different runs.}
\label{fig:MCstep_average}
\end{center}
\end{figure}

To estimate the efficiency of the present method, we first measure the average 
number $\bar n$ of potentials per site that survive as ${\bar V}_{ij}$
for several temperatures and $D$'s. The result is shown 
in Fig.~\ref{fig:Psurvive}. The average number $\bar n$ increases with decreasing temperature 
and increasing $D$. However, it is about $7$ even when $T=0.7~J$ and $D=0.1~J$. 
This means that more than $99.8\%$ of the interactions are cut off by being 
switched to $\tilde V_{ij}=0$. 
We next examine how $t_{\rm MC}$ depends on temperature, 
where $t_{\rm MC}$ is the total number of MC steps until $\Delta F$ is halved 20 times 
and the simulation is stopped. 
Figure~\ref{fig:MCstep_average} shows the result. 
The average is taken over 50 different runs. 
We measure $t_{\rm MC}$ for $D/J=0$ and $D/J=0.04$. 
We do not use the SCO method in the former case because long-range dipolar interactions are absent. 
In both cases, $t_{\rm MC}$ increases with decreasing temperature since relaxation becomes slower 
at lower temperatures. $t_{\rm MC}$ for $D/J=0.04$ increases more rapidly than that for $D/J=0$. 
However, the temperature dependence is not so strong. 
We also find that $t_{\rm MC}$'s for $D/J=0$ and $D/J=0.04$ are not 
very different. This result shows that $t_{\rm MC}$ does not increase much by the use of the SCO method. 
The computational time per run for $T=0.7~J$ and $D=0.1~J$, which is the most time-consuming case 
we have examined, was about ten days by a single-core calculation with a Core-i7 2.8~GHz processor. 

Finally, we again compare the present method with the conventional Wang-Landau method 
in which the joint density of states defined by eq.~(\ref{eqn:jointDOS}) is evaluated. 
The first merit of the present method is that the computational time for long-range 
interactions is markedly reduced by the use of 
the variant of the Wang-Landau method and the SCO method. 
As mentioned before, the difficulty in the conventional Wang-Landau method 
in long-range interacting systems is that we have to calculate the Hamiltonian 
${\cal H}\{ \vtr{S}_i \}$ after each update of a single spin. 
It is very desirable to develop a method of combining the conventional Wang-Landau 
method with the SCO method. To this end, an approach used in ref.~\citen{Sasaki10} 
might be helpful. The second merit of the present method is that the function 
$G(m_z)$ to be adjusted is a one variable function. In contrast, 
we have to adjust a two-variable function $G(E,m_z)$ in the conventional Wang-Landau method.
This is the main reason why the system size accessible by the present method 
is larger than that by the conventional Wang-Landau method~\cite{Zhou06}. 
However, the trade-off for this merit is that the temperature is kept constant. Therefore, 
in the present method, we can only estimate 
free energy at one temperature by a single simulation. In contrast, 
we can estimate free energy at any temperatures by a single simulation of the conventional Wang-Landau method 
because free energy at any temperatures can be calculated from 
the joint density of states using eq.~(\ref{eqn:FEintegral}). 
Furthermore, the present method has another drawback 
when the simulation is performed at low temperatures. In the conventional Wang-Landau method, 
high-energy states with high entropies are the source of fast relaxation, and 
the system rapidly forgets the present state when the system reaches a high-energy region. 
However, no such source exists in the present method when the temperature is low. 
Note that, in the present method, the zero-magnetization state is not 
a high-entropy state. When the temperature is low, the present method samples 
only a small portion of the states with low energies at any magnetization. This means that 
it is not trivial in the present method that equilibrium sampling is realized at low temperatures. 
Therefore, we should carefully check whether or not equilibrium sampling is realized. 
As performed in the present work, an effective way to check equilibration is by measuring free energy 
several times using different initial conditions and random sequences and by checking whether or not 
the same result is obtained. One should keep in mind that the present method 
has these drawbacks. 

In summary, we have developed an efficient MC method of free energy calculation 
in long-range interacting systems by combining a variant of the Wang-Landau method 
with the stochastic cutoff method. The efficiency of the method has been confirmed 
by applying the method to a free energy calculation in a three-dimensional magnetic dipolar system. 
We have also discussed the merits and demerits of the present method in comparison 
with the conventional Wang-Landau method.

The authors would like to thank Professor K. Sasaki for valuable discussions and comments. 
This work is supported by a Grant-in-Aid for Scientific Research (No. 21740279) from MEXT.








\end{document}